\documentclass[superscriptaddress,nofootinbib,amsmath,amssymb,aps,pra,twocolumn]{revtex4-2}

\usepackage{amsmath}
\usepackage{amssymb}
\usepackage{amsfonts}
\usepackage{bm} 
\usepackage{bbm}
\usepackage{braket}
\usepackage{color}
\usepackage{comment}
\usepackage{dcolumn} 
\usepackage{dsfont}
\usepackage{enumerate}
\usepackage{epsfig}
\usepackage{esint}
\usepackage[T1]{fontenc}
\usepackage{framed}
\usepackage{gensymb}
\usepackage{graphicx} 
\usepackage[colorlinks,linkcolor=blue,citecolor=blue,urlcolor=blue,hyperindex,driverfallback=dvipdfm]{hyperref}
\usepackage{indentfirst}
\usepackage{lmodern}
\usepackage{mathrsfs}
\usepackage{mathtools}
\usepackage{multirow}
\usepackage{psfrag}
\usepackage{pst-all}
\usepackage{soul}
\usepackage{xcolor}
\usepackage{xspace}
\usepackage{orcidlink}

\newcommand{\ccpar}[1] {\mathopen{}\left(#1\right)\mathclose{}}
\newcommand{\sqpar}[1] {\mathopen{}\left[#1\right]\mathclose{}}

\def\ii{{\rm i}}  \def\ee{{\rm e}}
  \def\kB{{k_{\rm B}}}
\def\Ree{{\rm Re}}  \def\Imm{{\rm Im}}

\def\xx{\hat{\bf x}}    
    
\def\Eb{{\bf E}}  
\def\EF{{E_{\rm F}}}  \def\vF{v_{\rm F}}  
\def\eps{\epsilon}  \def\vep{\varepsilon}
\def\ww{\omega}
\def\Hm{\mathcal{H}}
\def\phiext{\phi^{\rm ext}}  

\begin{document}

\title{Nonlinear thermoplasmonics in graphene nanostructures}

\author{Line Jelver\,\orcidlink{0000-0001-5503-5604}}
\affiliation{POLIMA---Center for Polariton-driven Light--Matter Interactions, University of Southern Denmark, Campusvej 55, DK-5230 Odense M, Denmark}

\author{Joel~D.~Cox\,\orcidlink{0000-0002-5954-6038}}
\email[Joel~D.~Cox: ]{cox@mci.sdu.dk}
\affiliation{POLIMA---Center for Polariton-driven Light--Matter Interactions, University of Southern Denmark, Campusvej 55, DK-5230 Odense M, Denmark}
\affiliation{Danish Institute for Advanced Study, University of Southern Denmark, Campusvej 55, DK-5230 Odense M, Denmark}

\begin{abstract}
The linear electronic dispersion relation of graphene endows the atomically thin carbon layer with a large intrinsic optical nonlinearity, with regard to both parametric and photothermal processes. While plasmons in graphene nanostructures can further enhance nonlinear optical phenomena, boosting resonances to the technologically relevant mid- and near-infrared (IR) spectral regime necessitates patterning on $\sim10$\,nm length scales, for which quantum finite-size effects play a crucial role. Here we show that thermoplasmons in narrow graphene nanoribbons can be activated at mid- and near-IR frequencies with moderate absorbed energy density, and furthermore can drive substantial third-harmonic generation and optical Kerr nonlinearities. Our findings suggest that photothermal excitation by ultrashort optical pulses offers a promising approach to enable nonlinear plasmonic phenomena in nanostructured graphene that avoids potentially invasive electrical gating schemes and excessive charge carrier doping levels.
\end{abstract}

\date{\today}
\maketitle

\section{Introduction}

Nonlinear optical phenomena facilitate the temporal and spectral control of light by light, a crucial resource for emerging integrated photonic and quantum optical technologies \cite{chang2014quantum,dutt2024nonlinear}. Despite the inherently weak nonlinear response of conventional materials \cite{boyd2020nonlinear}, demand for efficient miniaturized photonic devices motivates intensive research on material platforms and schemes to enable nonlinear light-matter interactions on the nanoscale \cite{kauranen2012nonlinear,li2017nonlinear,taghizadeh2021twodimensional}. Graphene is actively explored in this context due to its conical electronic dispersion relation, which endows the atomically thin carbon layer with a proportionately large nonlinear optical response \cite{mikhailov2007nonlinear}. Measurements of third-harmonic generation (THG), four-wave mixing, and the optical Kerr effect in graphene reveal large third-order susceptibilities, on the order of $10^{-16}-10^{-15}$\,m$^2$/V$^2$, several orders of magnitude larger than in other highly nonlinear optical materials \cite{hendry2010coherent,kumar2013third,dremetsika2017enhanced,jiang2018gatetunable,soavi2018broadband}.

In electrically doped graphene, the hybridization of light with plasmons---collective oscillations of conduction electrons---can strongly concentrate electromagnetic near fields to boost nonlinear optical effects. In particular, the combined linear electronic bands and atomic thickness of graphene results in plasmon resonances with high sensitivity to charge carrier doping levels \cite{garciadeabajo2014graphene}, enabling electrical tunability of plasmon resonances in the terahertz (THz) or infrared (IR) spectral bands \cite{gonccalves2016introduction}. Patterning of graphene into ribbons or islands offers the means to couple light with localized plasmon resonances that can be tuned passively through spatial confinement while maintaining active tunability via electrostatic gating \cite{brar2013highly}. 
While enhancements in the nonlinear response associated with THz plasmons in graphene nanoribbons (GNRs) have been reported in wave-mixing experiments \cite{kundys2018nonlinear}, large optical nonlinearities driven by plasmons at near-IR and visible frequencies in smaller graphene nanostructures, predicted to exhibit an involved dependence on quantum finite-size and edge termination effects \cite{cox2019nonlinear}, have yet to be observed. In this context, narrow GNRs of $\lesssim10$\,nm width constitute an appealing platform for explorations in nonlinear graphene plasmonics, particularly with recent advances in nanofabrication techniques enabling atomically-precise control of GNR width and edge configuration \cite{kimouche2015ultranarrow,jacobberger2015direct,zhao2020optical,lyu2024graphene}.

The linear electronic bands in graphene are not only associated with a large parametric nonlinear optical response, but also a low electronic heat capacity \cite{balandin2011thermal,pop2012thermal}, thus presenting opportunities to explore nonlinear dynamics associated with thermo-optical effects \cite{garciadeabajo2014graphene,hafez2020terahertz}. Energy absorbed by pumping graphene with intense ultrashort optical pulses is rapidly converted to electronic heat within $\sim10$s of femtoseconds \cite{tielrooij2013photoexcitation,mics2015thermodynamic,tomadin2018ultrafast}, such that the two-dimensional electron gas can reach temperatures of several $\sim1000$\,K using pulses of moderate $\lesssim1$\,J/m$^2$ optical fluence \cite{lui2010ultrafast}. Time- and angle-resolved photoemission spectroscopy measurements reveal that the thermalized carrier distribution relaxes on $\sim$picosecond timescales by exciting phonons in the carbon lattice \cite{johannsen2013direct,gierz2013snapshots}. The collective thermal response of graphene under illumination by THz pulses---for which electronic heating is effectively instantaneous relative to oscillations of the electromagnetic field---results in efficient harmonic generation, with third-harmonic susceptibilities reaching $\sim10^{-9}$ m$^2$/V$^2$ \cite{hafez2018extremely}. At IR frequencies, nonequilibrium intra-pulse dynamics in graphene can manifest as spectral blueshifts in third- and fifth-harmonic signals \cite{baudisch2018ultrafast}, while carrier heating effects have been found to significantly impact wave mixing and harmonic generation \cite{constant2016alloptical,soavi2018broadband}.

Electron heating in photoexcited graphene facilitates optical modulation of graphene plasmon resonances on ultrafast timescales \cite{vafek2006thermoplasma,dias2020thermal}. For example, experiments on graphene encapsulated in hBN show that optically-induced carrier heating can sustain mid-IR plasmons that are otherwise absent in the equilibrium state \cite{ni2016ultrafast}, while pump-probe measurements of THz plasmons in doped GNRs reveal a transient redshift in localized plasmon resonances \cite{jadidi2016nonlinear}. Qualitatively similar spectral shifts and saturation of mid-IR plasmons under resonant excitation are expected for narrower GNRs of a few $10$s of nm width \cite{cox2018transient}, while the distribution of absorbed optical energy among fewer free charge carriers in even smaller graphene nanostructures is predicted to enable more dramatic temperature-induced changes in the optical response \cite{cox2019single}.

In this letter, we explore possibilities to thermally induce mid- and near-IR plasmons in GNRs with $\lesssim10$\,nm width that can trigger a large nonlinear response. Combined ab-initio modelling and self-consistent perturbation theory reveals that quantum finite-size effects in GNRs significantly impact their supported thermoplasmons, particularly in the nonlinear response associated with third-harmonic generation. Prominent thermoplasmonic features are predicted to emerge in the linear and nonlinear spectral response after doping GNRs with moderate heat energy densities on the order of $\sim 1$\,J/m$^2$, below the fluences of optical pulses commonly used in ultrafast experiments. Our results indicate that optical pumping constitutes a viable approach to realizing a sizeable transient nonlinear plasmonic response at near-IR frequencies in mesoscopic graphene nanostructures, circumventing the need for cumbersome charge carrier doping schemes.

\begin{figure*}[t]
    \centering
    \includegraphics[width=1\textwidth]{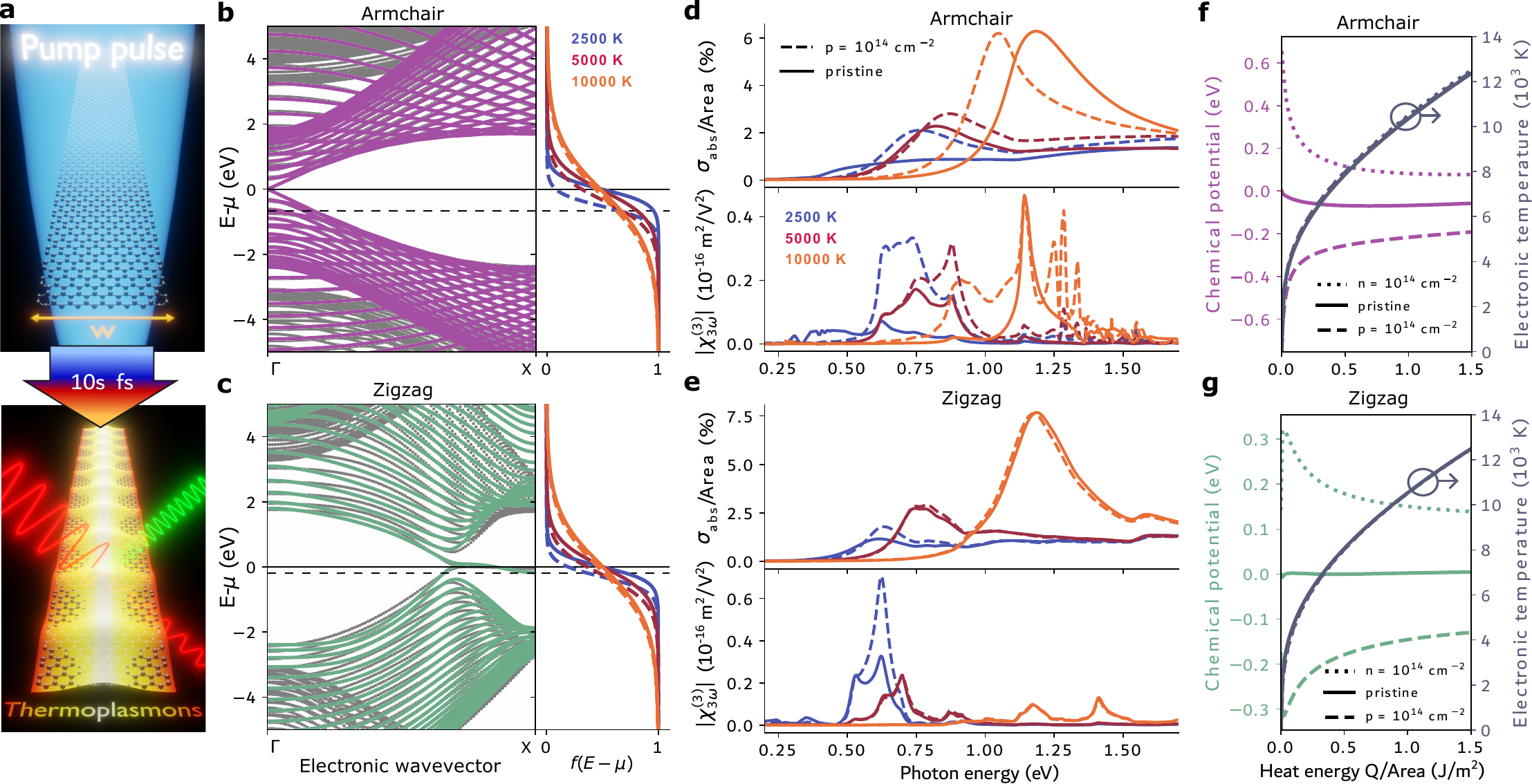}
    \caption{\textbf{Transient thermoplasmons in narrow graphene nanoribbons.} \textbf{a} Schematic illustration of a graphene nanoribbon (GNR) of width $W$ that is photoexcited (upper panel) by an intense ultrashort pulse, so that after a few $10$s of femtoseconds its electrons occupy a thermalized state that can be probed in nonlinear spectroscopy (lower panel) within $\sim1$ picosecond. Panels \textbf{b} and \textbf{c} show the electronic band structure (left) and occupation functions (right) of undoped armchair (AC) and zigzag (ZZ) edge-terminated GNRs, respectively, each of $W\approx 5$\,nm width, with occupation functions determined by Fermi-Dirac statistics at the indicated electron temperatures and corresponding chemical potentials satisfying Eq.~\eqref{eq:number}. The associated linear absorption cross section (upper panels) and third-harmonic susceptibility (lower panels) are shown in panels \textbf{d} and \textbf{e}, both for GNRs that are undoped (solid curves) and hole-doped to density $p=10^{14}$\,cm$^{-2}$ (dashed curves). Panels \textbf{f} and \textbf{g} show the electronic temperature and chemical potential determined from Eqs.~\eqref{eq:number} and \eqref{eq:Q} as functions of the added thermal energy $Q$ per unit area for the same $W\approx 5$\,nm AC and ZZ GNRs at high electron (dotted curves) and hole (dashed curves) doping levels, contrasted with undoped GNRs (solid curves).}
    \label{fig:fig1}
\end{figure*}

\section{Results and discussion}

\begin{figure*}[t]
    \centering
    \includegraphics[width=1\textwidth]{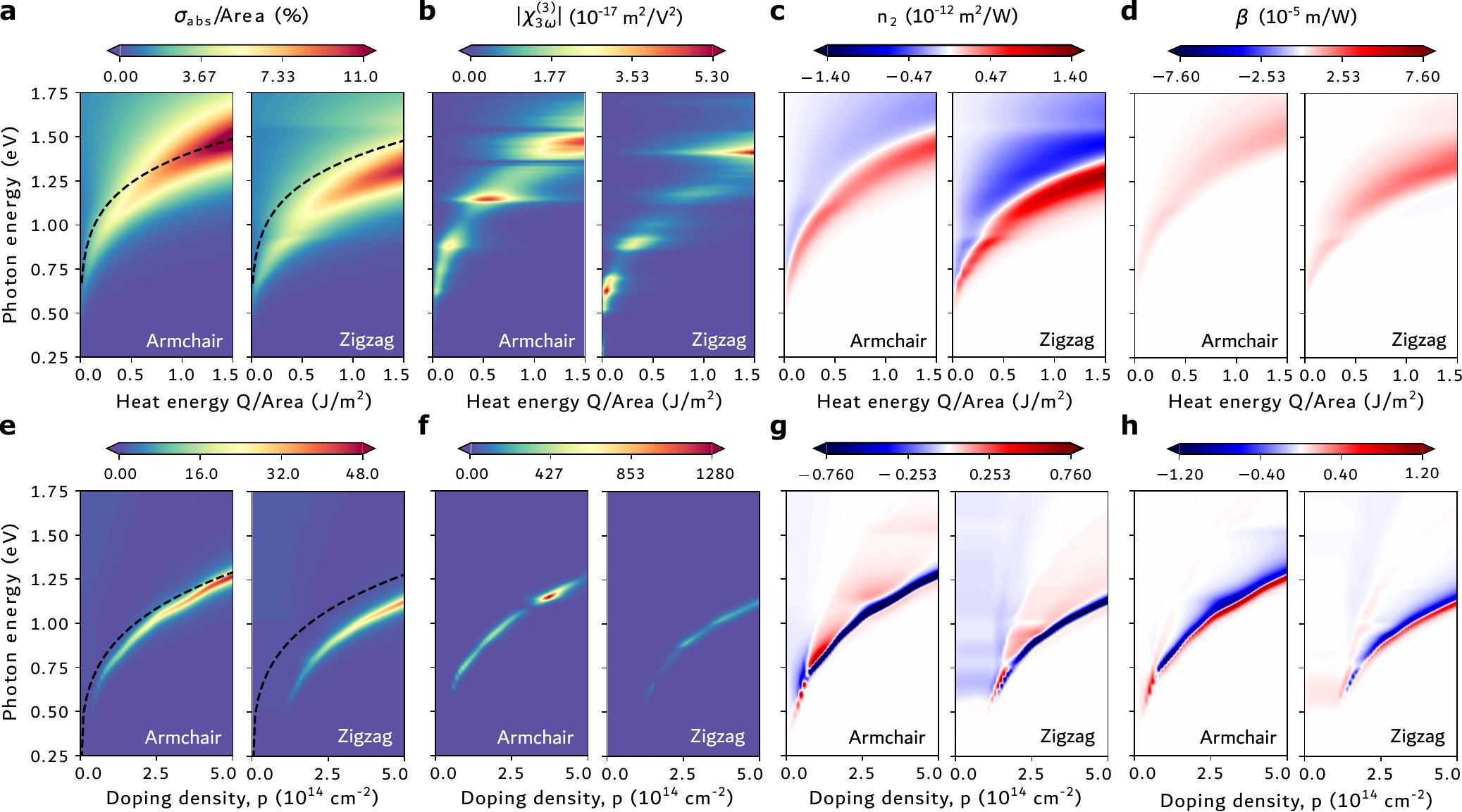}
    \caption{\textbf{Tunability of graphene plasmons with temperature and charge carrier doping.} The linear absorption cross section $\sigma^{\rm abs}$, third-harmonic susceptibility $\chi_{3\ww}^{(3)}$, nonlinear refractive index $n_2$, and two-photon absorption $\beta$ spectra of the $W\approx 5$\,nm GNRs considered in Fig.~\ref{fig:fig1} are presented as functions of \textbf{a} the added heat energy $Q$ per unit area and \textbf{b} the hole doping density $p$. Dashed lines in the $\sigma^{\rm abs}$ plots indicate the plasmon resonance energy predicted by the classical model of Eq.~\eqref{eq:classical}, while the black dots in \textbf{a} trace the frequency of maximal absorption near interband hybridization points indicated by arrows.}
    \label{fig:fig2}
\end{figure*}

Electrons in a narrow GNR that are photoexcited by an ultrafast pump pulse with moderate fluence will rapidly thermalize to temperatures on the order of $\sim1000$s\,K, effectively modifying the free charge carrier density within $\sim1$\,ps as the electrons transfer their energy to the carbon lattice. As illustrated schematically in Fig.~\ref{fig:fig1}a, transient plasmon resonances sustained by the thermalized electron distribution can be optically probed in both linear and nonlinear optical response regimes. Electronic states in armchair (AC) and zigzag (ZZ) edge-terminated GNRs of $\sim5$\,nm width obtained from first-principles density functional theory (DFT) calculations are plotted in Figs.~\ref{fig:fig1}b and \ref{fig:fig1}c, respectively, as functions of the electronic wave vector in the direction of translational invariance (see Methods). The population of a state with energy $\hbar\vep_{j,k}$ indexed by band $j$ and wave vector $k$ is determined from Fermi-Dirac statistics as
\begin{equation} \label{eq:FD}
    f_{j,k}(\mu, T) = \left[\ee^{(\hbar\vep_{j,k}-\mu)/\kB T}+1\right]^{-1}
\end{equation}
for a chemical potential $\mu$ and electron temperature $T$ satisfying the particle number conservation condition
\begin{equation} \label{eq:number}
    \frac{b}{2\pi}\int_{-\pi/b}^{\pi/b}{\rm d}k\sum_j\sqpar{f_{j,k}(\mu,T)-f_{j,k}(\EF,0)} = 0 ,
\end{equation}
where the integration over $k$ is defined by the GNR unit cell period $b$ and $\EF$ is the Fermi energy. The panels plotted alongside the electronic bands in Fig.~\ref{fig:fig1}b and c show the corresponding Fermi-Dirac distribution obtained for pristine GNRs, i.e., when $\EF=0$ (solid curves), as well as for hole-(p-)doped GNRs at doping density $p=10^{14}$\,cm$^{-2}$ (dashed curves), at the specified electron temperatures and corresponding chemical potentials that satisfy Eq.~\eqref{eq:number}.

To study the optical properties of thermalized GNRs, we apply the ab-initio framework reported in Ref.~\cite{jelver2023nonlinear} to construct tight-binding Hamiltonians for carbon p$_z$ orbitals using maximally-localized Wannier functions (MLWFs). As shown in Figs.~\ref{fig:fig1}b and \ref{fig:fig1}c, the Wannier tight-binding (WTB) models accurately reproduce the ab-initio bands corresponding to p$_z$ orbitals at low energies around the Fermi level that are most relevant for electron temperatures $T\lesssim10000$\,K, as indicated by the distributions plotted alongside the electronic bands. The obtained WTB Hamiltonians are used to compute the linear and nonlinear optical response of GNRs according to a perturbative solution of the equation of motion governing the single-particle density matrix \cite{cox2016quantum}. Further details on the ab-initio electronic structure calculations and optical response simulations are provided in Methods, while additional calculations for narrower and wider ribbons are provided in Supplementary Material (SM).

In Figs.~\ref{fig:fig1}d and \ref{fig:fig1}e, we present the linear absorption cross-section and the nonlinear susceptibility associated with THG computed for the AC and ZZ GNRs considered in Figs.~\ref{fig:fig1}b and \ref{fig:fig1}c, respectively, under the same electronic temperature and charge carrier doping conditions. Temperature-induced plasmon resonances appear in the linear response already at $T=2500$\,K, which are slightly more pronounced in ZZ GNRs, while the thermoplasmons blueshift and become more prominent as the electron temperature is increased to $10000$\,K. Comparison of the pristine and p-doped ribbons reveals relatively minor changes in the linear response associated with thermoplasmons due to charge carrier injection, indicating that the temperature smearing of the electron distribution plays a much more important role than the doping density. Interestingly, an analysis of smaller GNRs presented in the SM reveals that the suggested crossover from charge doping-induced plasmons to thermoplasmons consistently occurs at relatively low temperature. The THG response shown in the lower panels of Figs.~\ref{fig:fig1}d and \ref{fig:fig1}e shows strikingly different behavior, e.g., a more noticeable dependence on the charge carrier doping, while the thermoplasmonic THG response of the ZZ GNR tends to decrease with increasing electronic temperature.

The large electronic temperatures required by thermoplasmons in GNRs are associated with thermal energy $Q$ that can be delivered by ultrashort optical pulses. From the electronic band structures computed for the one-dimensional GNRs considered here, we determine the chemical potential $\mu$ and electronic temperature $T$ by simultaneously imposing the particle number conservation condition in Eq.~\eqref{eq:number} and energy conservation
\begin{equation} \label{eq:Q}
    \frac{b}{2\pi}\int_{-\pi/b}^{\pi/b}{\rm d}k\sum_j \hbar\vep_j\left[f_{j,k}(\mu,T)-f_{j,k}(\EF,0)\right] = Q .
\end{equation}
The evolution of $\mu$ and $T$ with the heat energy per unit area $Q/{\rm Area}$ is presented in Figs.~\ref{fig:fig1}f and \ref{fig:fig1}g for AC and ZZ ribbons, respectively, in pristine (solid curves), electron-doped (density $n=10^{14}$\,cm$^{-2}$, dotted curves), and hole-doped (density $p=10^{14}$\,cm$^{-2}$, dashed curves) specimens. Independently of the considered charge doping conditions, electronic temperatures of several $1000$\,K are attained for heat energy levels commensurate with pulse fluences used in ultrafast optical experiments \cite{lui2010ultrafast}.

In Fig.~\ref{fig:fig2} we study the linear and nonlinear optical response of the $W\approx 5$\,nm AC and ZZ GNRs considered in Fig.~\ref{fig:fig1} when either thermal energy (Fig.~\ref{fig:fig1}a) or additional charge carriers (Fig.~\ref{fig:fig1}b) are added. Thermoplasmon resonances, appearing as prominent features in the linear absorption cross sections of Fig.~\ref{fig:fig2}a that blueshift with increasing heat energy, are relatively well-reproduced by the dashed curve, obtained from the combined intraband conductivity of extended graphene and classical electrodynamic model of plasmon resonances in GNRs as
\begin{equation} \label{eq:classical}
    \omega_{\rm p} = \frac{e/\hbar}{\sqrt{-\pi\eta_1\epsilon}}\sqrt{\frac{\mu_{\rm D}}{W}} ,
\end{equation}
where $\mu_{\rm D}=\mu + 2\kB T\log\ccpar{1+\ee^{-\mu/\kB T}}$ is the temperature-dependent Drude weight, $\eps=1$ is the relative permittivity of the environment, and $\eta_1=-0.06873$ is the eigenvalue associated with the leading dipolar plasmon resonance supported by the ribbon geometry \cite{yu2017analytical}. Specifically, the dashed curve in Fig.~\ref{fig:fig2}a is given from Eq.~\eqref{eq:classical} by setting $\mu=\EF=0$ and using the electron temperature obtained from the heat capacity of extended graphene, $Q/{\rm Area}=\beta (k_{\rm B}T)^3/(\hbar\vF)^2$, where $\beta\approx 2.2958$ in the $\EF\ll\kB T$ limit considered here \cite{yu2017ultrafast}. The systematic overestimation of the thermoplasmon resonance energy in the classical model is attributed here to its omission of Landau damping by interband transitions, which, in the case of narrow GNRs, can become more prominent at specific frequencies due to the large energy gaps between electronic bands associated with quantum confinement. In particular, we find that thermoplasmons can strongly hybridize with single-particle transitions at fixed frequencies to create Fano-like spectral features in the absorption cross section. Incidentally, for the same added heat energy, the thermoplasmon energy in ZZ GNRs is reduced compared to their AC counterparts due to the presence of edge states.

\begin{figure*}
    \centering
    \includegraphics[width=1\textwidth]{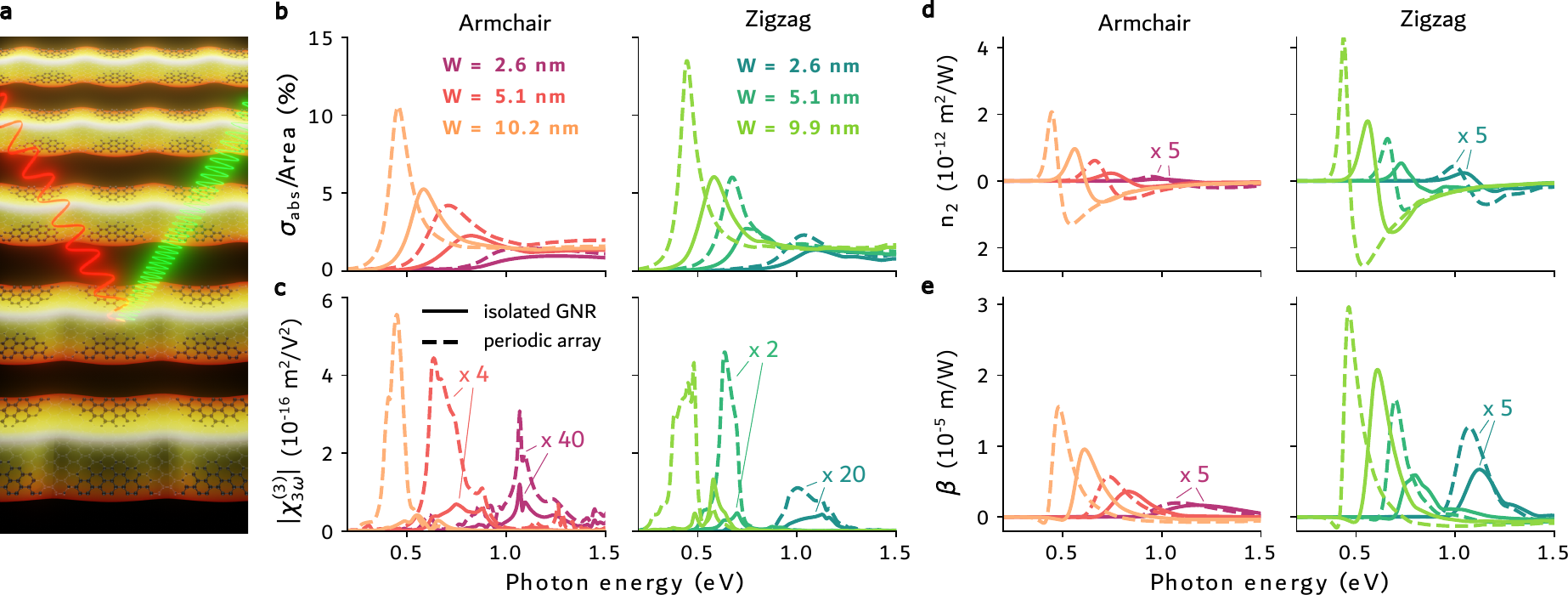}
    \caption{\textbf{Thermoplasmons in graphene nanoribbon arrays.} \textbf{a} Illustration of enhanced third-harmonic generation by thermoplasmons in a GNR array. The spectral response associated with \textbf{b} linear absorption, \textbf{c} THG, \textbf{d} the nonlinear refractive index, and \textbf{e} the two-photon absorption coefficient are shown for armchair (left columns) and zigzag (right columns) edge-terminated ribbons, with solid and dashed curves indicating the response of isolated $W\approx 5$\,nm GNRs and their periodic arrays with 1\,nm edge separation, respectively.}
    \label{fig:fig3}
\end{figure*}

Quantum finite-size effects are found to manifest more clearly in the thermoplasmon-driven THG spectra shown in Fig.~\ref{fig:fig2}a, which are not only sensitive to the superlinear dependence on plasmon field enhancement at the fundamental frequency that scales as $\sim\sqrt{\mu_{\rm D}}$, but also to single-particle transitions at higher harmonics. The intersection of these effects results in regions of pronounced enhancement and quenching in the THG spectra that are independent of the added heat energy. These findings suggest that, at particular frequencies, photoexcited transient thermoplasmons in narrow GNRs should produce a relatively large THG response over a longer duration as heat energy is dissipated to phonons in the carbon lattice. Qualitatively different behavior is observed in the optical Kerr effect, quantified by the intensity-dependent nonlinear refractive index $n_2$ and the two-photon absorption coefficient $\beta$, respectively, which evolve monotonically with the thermoplasmon feature in the linear absorption spectra. Notably, the heat-independent hybridization features seen in the linear response are more clearly resolved in the Kerr nonlinearity, for which the thermoplasmon-enhanced near fields at the driving and generated frequencies coincide. The behavior described above is corroborated by additional simulations of thermoplasmons in narrower and wider GNRs presented in the SM: in narrower GNRs, the thermoplasmonic response exhibits larger spectral ``jumps'' with varying heat energy, owing to the greater energetic spacing between electronic bands associated with quantum confinement effects; conversely, the reduced energy spacing of electronic bands in wider GNRs leads to a smoother spectral evolution of thermoplasmons with heat energy. 

Compared to the thermoplasmons activated with $\lesssim1$\,J/m$^2$ heat energy, the plasmonic resonances supported by electrically-doped GNRs at room temperature appear as sharper spectral features in Fig.~\ref{fig:fig2}b, however requiring doping densities $p\gtrsim10^{14}$\,cm$^{-2}$ that may be difficult to achieve through electrostatic gating. In the linear absorption spectra, plasmon energies are similarly well described by the classical model of Eq.~\eqref{eq:classical} taking $\mu=\hbar\vF\sqrt{\pi p}$, where $\vF\approx c/300$ is the Fermi velocity in graphene, albeit overestimating the plasmon resonance frequencies in the ZZ GNR owing to the presence of edge states. While the large absorption cross sections at plasmon resonances correspond to higher in-plane electric field intensities that lead to a proportionately larger THG response, the effective Kerr nonlinearity driven by doping charge carrier plasmons is actually reduced compared to that associated with thermoplasmons. We attribute this phenomenon to the interplay between enhanced Ohmic losses at the plasmon resonance and the sensitivity of the intrinsic Kerr nonlinearity to the electronic distribution in the GNRs, both of which are accounted for in the definitions of $n_2$ and $\beta$ (see Methods). Interestingly, the thermoplasmon-driven two-photon absorption coefficient is mainly positive, while the doping charge plasmon produces both positive and negative values of $\beta$, corresponding to TPA and saturable absorption, respectively.

We have thus far focused on the thermo-optical response of individual GNRs, while realistic pump-probe experiments should involve ribbon arrays to accumulate detectable far field signals from transient thermoplasmons \cite{garciadeabajo2014graphene,jadidi2016nonlinear}. The fabrication of narrow GNR arrays, with nanometer-scale control over ribbon width and spacing \cite{wang2021graphene}, presents opportunities for engineering tunable plasmonic devices that exploit plasmon hybridization effects \cite{silveiro2015quantum,semenenko2018plasmonplasmon,siegel2021using}. In Fig.~\ref{fig:fig3}, we study the linear and nonlinear optical response of $T=5000$\,K thermoplasmons in undoped isolated AC and ZZ GNRs of selected widths $W$ and their arrays at a fixed 1\,nm edge separation. The enhanced filling factors in arrays of wider ribbons lead to larger plasmon redshifts and absorption cross sections, while also boosting the associated THG and Kerr nonlinearity.

In conclusion, we have shown that narrow GNRs of $\lesssim10$\,nm width can support localized thermoplasmons at near-IR frequencies that enhance parametric nonlinear light-matter interactions. Thermal doping of narrow GNRs can be further exploited to study quantum confinement effects, both in the linear and nonlinear optical response regimes, with third-harmonic generation exhibiting a particularly large sensitivity to electronic structure. Our results indicate that thermoplasmons can be activated with moderate added heat energy, presenting opportunities to optically induce transient near-IR plasmons in small graphene nanostructures, without resorting to complicated electrical doping setups.

\section{Methods}

The electronic groundstates of one-dimensional AC and ZZ GNRs are computed using density functional theory, specifically using the Perdew-Berke-Ernzhofer functional as implemented in the Quantum Espresso code \cite{perdew1996generalized,giannozzi2009quantum,giannozzi2017advanced}. Wavefunctions are expanded in plane waves using normconserving pseudo potentials from the PseudoDojo database \cite{vansetten2018pseudodojo}, with a cut-off energy of 86\,Ry. We use 12 and 24 k-point grids in the direction of translational invariance for AC and ZZ ribbons, respectively, and a Gamma point sampling in the two directions across ribbons. Each ribbon is edge-passivated by hydrogen atoms, while the interatomic forces and cell stress are minimized. The isolated ribbons are separated by 3\,nm in both lateral directions to avoid interactions between neighbouring images. The wannierization of the carbon p$_z$ orbitals and creation of the Wannier tight-binding (WTB) Hamiltonians $\Hm^{\rm TB}$ are performed using the Wannier90 code \cite{pizzi2020wannier90}, from which the energy $\hbar\vep_{j,k}$ of band $j$ at Bloch wave vector $k$ is determined by diagonalizing
\begin{equation}
    \sum_{l'}\Hm_{ll',k}^{\rm TB}a_{jl',k} = \hbar\vep_{j,k}a_{jl,k} ,
\end{equation}
where $a_{jl,k}$ are expansion coefficients that weigh the contribution of orbital $l$ to the ribbon eigenstates.

Following the approach of Refs.~\cite{cox2016quantum,jelver2023nonlinear}, the WTB eigenstates are used in the framework of the random phase approximation (RPA) to construct the non-interacting susceptibility 
\begin{gather} \label{eq:xi0}
    \chi_{s\ww,ll'}^0 = \frac{e^2b}{\pi\hbar}\int_{-\pi/b}^{\pi/b}{\rm d}k\sum_{jj'}\ccpar{f_{j',k}-f_{j,k}} 
 \\
 \times\frac{a_{jl,k}a_{j'l,k}^*a_{jl',k}^*a_{j'l',k}}{\ww+\ii/2\tau-\ccpar{\vep_{j,k}-\vep_{j',k}}}  \nonumber
\end{gather}
that determines the induced charge density component
\begin{equation}
    \rho_l^{ns} = \sum_{l'}\chi_{s\ww,ll'}^{0}\phi_{l'}^{ns} + \rho_{{\rm NL},l}^{ns}
\end{equation}
quantifying the response at order $n>0$ oscillating with harmonic $s$ ($|s|<n$) of the external field $\Eb^{\rm ext}\ee^{-\ii\ww t}+{\rm c.c.}$ in the perturbation series $\rho_l(t) = \sum_{n,s}\rho_l^{ns}\ee^{-\ii s\ww t}$, where
\begin{equation}
    \phi_l^{ns} = \phiext_{s\ww,l}\delta_{n,1} + \sum_{l'}v_{ll'}\rho_{l'}^{ns}
\end{equation}
is the self-consistent potential, accounting for the external potential $\phiext_{s\ww,l}$ at linear order ($n=1$) and the Coulomb interaction $v_{ll'}$ between sites $l$ and $l'$, while $\rho_{{\rm NL},l}^{ns}$ is a nonlinear source term (nonvanishing for $n>1$) constructed from lower perturbation orders. In Eq.~\eqref{eq:xi0}, the dependence on chemical potential and electron temperature enters the occupation factors $f_{j,k}$ according to Eq.~\eqref{eq:FD}, while inelastic scattering processes are described by a phenomenological damping parameter that we set to a conservative value of $\hbar\tau^{-1}=50$\,meV throughout this work. Combining the above expressions, the density can be expressed as
\begin{equation}
    \rho^{ns} = \left(1-\chi_{s\ww}^{0}v\right)^{-1}\cdot\ccpar{\chi_{s\ww}^0\cdot\phiext_{s\ww}\delta_{n,1}+\rho_{\rm NL}^{ns}}
\end{equation}
in matrix notation. For normally-impinging fields polarized along the ribbon confinement direction $\xx$, i.e., $\phiext_{s\ww,l}=-x_lE^{\rm ext}$, we compute the polarizability per unit length $\alpha_{s\ww}^{(n)}/L=\sum_lx_l\rho_l^{ns}/(E^{\rm ext})^nb$, where $b$ is the unit cell length along the ribbon. In practice, we use the parametrized Coulomb interaction $v_{ll'}$ of carbon $p_z$ orbitals described in Ref.~\cite{thongrattanasiri2012quantum}, which is modified in the case of ribbon arrays to also include the interactions of site $l$ with the corresponding sites $l'$ in all other ribbons \cite{silveiro2015quantum}.

Here we define the linear absorption cross section from the polarizability as $\sigma^{\rm abs}=(4\pi\ww/c)\Imm\{\alpha_\ww^{(1)}\}$, while the nonlinear suceptibilities are calculated from $\chi^{(n)}_{s\ww}=\alpha_{s\ww}^{(n)}/LWd_{\rm gr}$ by estimating the graphene thickness $d_{\rm gr}=0.33$\,nm as the interlayer spacing in graphite. In absorbing media, the nonlinear refractive index $n_2$ and two-photon absorption coefficient $\beta$ are given as \cite{delCoso2004}
\begin{subequations} \label{eq:Kerr}
\begin{align}
    n_2 &= \frac{3}{4\epsilon_0 c (n_0^2 + k_0^2)} \left( \Ree\{\chi^{(3)}_{\omega}\} + \frac{k_0}{n_0} \Imm\{\chi^{(3)}_{\omega}\} \right) , \\
    \beta &= \frac{3\omega}{2\epsilon_0 c^2 (n_0^2 + k_0^2)} \left( \Imm\{\chi^{(3)}_{\omega}\} - \frac{k_0}{n_0} \Ree\{\chi^{(3)}_{\omega}\}\right) ,
\end{align}    
\end{subequations}
where $n_0=\Ree\{(1+\chi^{(1)}_{\omega})^{1/2}\}$ and $k_0=\Imm\{(1+\chi^{(1)}_{\omega})^{1/2}\}$ are the linear refractive and absorptive index, respectively, $\epsilon_0$ is the vacuum permittivity, and $c$ is the speed of light. 

\section{Acknowledgements}

The authors thank Eduardo~J.~C.~Dias for insightful and enjoyable discussions.
J.~D.~C. is a Sapere Aude research leader supported by Independent Research Fund Denmark (grant no. 0165-00051B). The Center for Polariton-driven Light--Matter Interactions (POLIMA) is funded by the Danish National Research Foundation (Project No.~DNRF165).
Part of the computation in this project was performed on the DeiC Large Memory HPC system managed by the eScience Center at the University of Southern Denmark.

\section{Author contributions}

J.~D.~C. conceived the project and developed the theory. L.~J. refined the idea and performed the numerical calculations. All authors contributed to the analysis of the results and writing of the manuscript. 


%

\end{document}